\documentclass[preprint]{elsarticle}

\usepackage{amstext}
\usepackage{amssymb}
\usepackage{graphicx}
\usepackage{latexsym}
\usepackage{mathrsfs}
\usepackage{calrsfs}
\usepackage{amsfonts}
\usepackage{hyphenat}

{\count255=\time\divide\count255 by 60 \xdef\hourmin{\number\count255}
  \multiply\count255 by-60\advance\count255 by\time
  \xdef\hourmin{\hourmin:\ifnum\count255<10 0\fi\the\count255}}

\def\rd{ {\rm d}}
\def\vev#1{ \left\langle #1 \right \rangle }

\def\code#1{{\textsc #1}}
\def\func#1{{\it #1}}
\def\file#1{{\tt #1}}

\begin{document}

\title{\code{Cosmo++}: An Object-Oriented C++ Library for Cosmology}

\author{Grigor Aslanyan}
\ead{g.aslanyan@auckland.ac.nz}
\address{Department of Physics, University of Auckland, Private Bag 92019, Auckland, New Zealand}

\begin{abstract}
This paper introduces a new publicly available numerical library for cosmology, \code{Cosmo++}. The library has been designed using object-oriented programming techniques, and fully implemented in C++. \code{Cosmo++} introduces a unified interface for using most of the frequently used numerical methods in cosmology. Most of the features are implemented in \code{Cosmo++} itself, while a part of the functionality is implemented by linking to other publicly available libraries. The most important features of the library are Cosmic Microwave Background anisotropies power spectrum and transfer function calculations, likelihood calculations, parameter space sampling tools, sky map simulations, and mask apodization. \code{Cosmo++} also includes a few mathematical tools that are frequently used in numerical research in cosmology and beyond. A few simple examples are included in \code{Cosmo++} to help the user understand the key features. The library has been fully tested, and we describe some of the important tests in this paper. \code{Cosmo++} is publicly available at {\it http://cosmo.grigoraslanyan.com}.
\end{abstract}


\maketitle

\section{Introduction}

Numerical methods are an indispensable part of modern research in cosmology, from theoretical studies to experimental data analysis. In this paper we introduce a new numerical library for cosmology in C++, \code{Cosmo++}. We use object-oriented programming to separate different parts of the library into different classes, resulting in an intuitive and easy-to-understand user interface.

Numerous excellent numerical libraries and packages have been developed and rigorously tested in the past 2 decades, including \code{HEALPix} \cite{Gorski:2004ku}, \code{CAMB} \cite{Lewis:1999ju}, \code{CLASS} \cite{Lesgourgues:2011ud,Blas:2011vf}, WMAP likelihood code \cite{Jarosik:2010tx,Larson:2010ji,Komatsu:2010in,2013ApJS..208...19H}, Planck likelihood code \cite{Collaboration:2013vc}, \code{CosmoMC} \cite{Lewis:2002hg}. \code{Cosmo++} is not intended to replace any of the existing packages, but rather complement them with a clear interface and some additional tools that are frequently used in cosmological research. The main features of the library are likelihood calculation for Cosmic Microwave Background (CMB) data, parameter space sampling tools, an interface to the publicly available Boltzmann code \code{CLASS} \cite{Lesgourgues:2011ud,Blas:2011vf} for CMB power spectrum and transfer function calculations, a mask apodization tool, sky simulation tools, and a few mathematical utilities frequently used in numerical research in physics. The library is made of many independent modules (classes in C++), which can be used on their own or easily combined together. This allows the user to include their own code if desired, without any need to modify the existing code. They can implement their own version of any of the modules and combine them with the rest of the library. This is accomplished through the usage of inheritance, polymorphism, and template parameters. For example, the primordial power spectrum is treated as just a regular real function. To implement a new form of the primordial power spectrum, the user needs to simply inherit a new class from the given abstract class of a real function, and then pass an object of their new class into the functions of the library.

Since there are a few publicly available excellent Boltzmann solvers  \cite{Lewis:1999ju,Lesgourgues:2011ud,Blas:2011vf}, \code{Cosmo++} does not include a new one. It rather includes a clear interface for a Boltzmann solver, which is linked to \code{CLASS}. The Boltzmann solver is included as a completely independent module, allowing the user to easily switch to a different one.

Likelihood calculation tools are included both for large and small scales. One important feature of \code{Cosmo++} is temperature and polarization likelihood calculation for off-diagonal covariance matrices in harmonic space. The standard $\Lambda$CDM model assumes statistical isotropy of the universe, implying diagonal covariance matrices in harmonic space. However, some proposed extensions of $\Lambda$CDM, such as a non-trivial topology of the universe \cite{Collaboration:2013wv,Aslanyan:2013lsa} and anisotropic models of inflation \cite{Kim:2013vk,Groeneboom:2009wj}, break the statistical isotropy of the universe. In light of the large-scale anomalies detected in the CMB data \cite{Collaboration:2013vj}, anisotropic models of the universe need to be rigorously tested. The large-scale likelihood calculators included in \code{Cosmo++} are indispensable tools for such tests.

Two different parameter space samplers are included in \code{Cosmo++}, a Metro\hyp{}polis-Hastings sampler \cite{mcmc} and an interface to the publicly available MultiNest sampler \cite{Feroz:2008eb,Feroz:2013uq,Feroz:2007fi}. This allows the user to easily switch between the samplers or even implement and include their own sampler.

The library is fully documented using \code{Doxygen}\footnote{http://www.doxygen.org}. In this paper we describe the general functionality of the library, and the specifics of implementation.

This paper is organized as follows. Sections \ref{params_sec} -  \ref{utilities_sec} describe the main functionality of the library. In Section \ref{tools_sec} we describe some mathematical tools that are frequently used in cosmology and other areas of physics. We introduce a few examples in Section \ref{examples_sec}, showing the usage of the main features of the library in only a few lines of code. We describe some of the most important tests performed in Section \ref{tests_sec}, and we summarize in Section \ref{summary_sec}. Our notation and units are described in \ref{notation_sec}.

\section{Cosmological Parameters and Power Spectra}\label{params_sec}

\subsection{Cosmological Parameters}

The cosmological parameters can be described by inheriting a class from the abstract class \func{CosmologicalParams} (defined in \file{cosmological\_params.hpp}) and implementing all of the virtual functions. The virtual functions must simply return the values of different parameters. For example, the function \func{getOmBH2} must return the value of $\Omega_bh^2$ (see \ref{cosmo_params_ap} for our notation).

The class includes all of the standard $\Lambda$CDM parameters, as well as certain frequently used non-standard parameters. Instances of this class can be used to pass the cosmological parameters to other modules of the library. In particular, the class \func{CMB} described below takes an object of type \func{CosmologicalParams} as its input to calculate the CMB power spectra and transfer functions.

The primordial power spectrum is defined in \func{CosmologicalParams} as a general real function, as described below in Section \ref{ps_sec}. In addition to that, \func{CosmologicalParams} includes the parameters describing the primordial power spectrum in the standard case, such as $A_s$ and $n_s$. These parameters contain redundant information, and are only included for completeness. The code uses the general form of the power spectrum by default, and ignores the values of the redundant parameters. However, some parts of the code, such as the class \func{CMB}, allow the usage of the parameters instead of the general form of the power spectrum.

Some specific examples of cosmological parameters classes are included in the file \file{cosmological\_params.hpp}. The class \func{LambdaCDMParams} implements the parameters for the standard $\Lambda$CDM model. \func{LCDMWithTensorParams} describes the cosmological parameters for tensor fluctuations in addition to the standard $\Lambda$CDM parameters. \func{LCDMWithDegenerateNeutrinosParams} includes massive neutrinos of the same mass.

\subsection{Primordial Power Spectrum}\label{ps_sec}

The primordial power spectrum is simply described as a real function $\Delta^2(k)$, both for the scalar and the tensor case (see \ref{primordial_pert_ap} for our notation). To implement a primordial power spectrum one needs to inherit a class from the \func{Math::RealFunction} class (defined in the file \file{function.hpp}) and implement the virtual function \func{evaluate}. Then instances of that class can be passed to the other modules of the library to describe the primordial power spectrum.

A few standard cases of the primordial power spectrum have been implemented in the file \file{power\_spectrum.hpp}. The \func{StandardPowerSpectrum} and \func{StandardPowerSpectrumTensor} classes implement the standard power spectra (\ref{scalar_ps}) and (\ref{tensor_ps}), respectively.

\func{LinearSplinePowerSpectrum} and \func{CubicSplinePowerSpectrum} classes implement the cases when the primordial power spectrum can be written as a linear or cubic spline of a given number of knots \cite{Bridges:2009dj,Verde:2008er,Peiris:2009ke,Aslanyan:2014mqa,Abazajian:2014vr}.

\subsection{CMB Power Spectra and Transfer Functions}

The CMB power spectra and transfer functions can be calculated using the class \func{CMB} in \file{cmb.hpp}. The underlying code for these calculations is the publicly available code \code{CLASS} \cite{Lesgourgues:2011ud,Blas:2011vf}. \code{Cosmo++} does not include a new Boltzmann solver, it only provides a different interface for using the functionality of \code{CLASS}. This interface makes it possible to easily link the functionality provided by \code{CLASS} with the rest of the tools of the library.

\func{CMB} has to be initialized in two stages. The function \func{preInitialize} is used to set the general parameters, such as the maximum value of $l$ to be used in calculations. After this the function \func{initialize} can be called to set the cosmological parameters, following which the functions \func{getCl}, \func{getLensedCl}, and \func{getTransfer} can be called to retrieve various CMB power spectra and transfer functions. In addition, the functions \func{getMatterPs} and \func{getMatterTransfer} will retrieve the matter power spectrum and transfer function at a given redshift, and \func{sigma8} will calculate the $\sigma_8$ value. In case one needs to calculate the CMB power spectra for different values of cosmological parameters, one can call the function \func{initialize} multiple times after one call of \func{preInitialize}. The division of initialization into two stages has been done with that scenario in mind; not repeating the pre-initialization saves some computing time.

The parameter \func{primordialInitialize} passed to the function \func{preInitialize} defines if the primordial power spectrum should be initialized from the general functional form, or simply from the standard parameters $A_s$, $n_s$ for the scalar case and $r$, $n_t$ for the tensor case. By default, the general functional form is used for primordial power spectra, however if one is using the standard case some computing time can be saved by setting the parameter \func{primordialInitialize} to \func{false} and using the standard parameters.

The parameter \func{wantAllL} defines if the power spectra and the transfer functions should be calculated for all values of $l$. By default, the calculation of the power spectra is done for only some values of $l$ and interpolated to get the full spectrum. This speeds up the calculation by about one order of magnitude with a negligible loss in accuracy. However, if very accurate calculations of the power spectra are needed, the parameter \func{wantAllL} needs to be set to \func{true}. In particular, if one needs the transfer functions by themselves, this parameter needs to be set to \func{true}.

\section{Likelihood Calculation}\label{likelihood_sec}

Likelihood calculation is a crucial step in any parameter estimation pipeline. \code{Cosmo++} provides three different tools for likelihood calculation, which can be used by themselves or in combination. The perturbations are assumed to be Gaussian for all three cases.

\subsection{Low-$l$ Likelihood}

In the low-$l$ range the distribution of the $C_l$'s cannot be well approximated by a Gaussian distribution \cite{Efstathiou:2003bh}. For this reason likelihood calculation for low multipoles is done in pixel space. \code{Cosmo++} includes pixel space likelihood calculation functionality for temperature and polarization maps.

For Gaussian perturbations the likelihood function takes the form
\begin{equation}
\mathcal{L}(\mathbf{m}|S)\rd\mathbf{m}=\frac{\exp\left[-\frac{1}{2}\mathbf{m}^t(S+N)^{-1}\mathbf{m}\right]}{(2\pi)^{3n_p/2}|S+N|^{1/2}}\rd\mathbf{m}
\end{equation}
where $\mathbf{m}=(\mathbf{T},\mathbf{Q},\mathbf{U})$; $S$ and $N$ are the signal and noise covariance matrices, respectively; $n_p$ is the number of pixels; and the symbol $|..|$ stands for determinant. The rest of the notation is explained in \ref{cmb_maps_ap}.

For low-$l$ modes the noise in the temperature can usually be ignored, in which case it is easier to decompose the likelihood function into temperature and polarization parts \cite{Page:2006hz}. We also assume no parity violation, implying $M^{TB}=0$ and $M^{EB}=0$. We then define
\begin{equation}\label{e_redef}
\tilde{E}_{lm}=E_{lm}-M^{TE}_{l^\prime m^\prime lm}(M^{TT}_{l^\prime m^\prime l^{\prime\prime}m^{\prime\prime}})^{-1}T_{l^{\prime\prime}m^{\prime\prime}}
\end{equation}
which gives
\begin{equation}
\vev{\tilde{E}_{lm}T_{l^\prime m^\prime}^*}=0\,,
\end{equation}
\begin{equation}\label{new_EE}
\vev{\tilde{E}_{lm}\tilde{E}_{l^\prime m^\prime}^*}\equiv M^{\tilde{E}\tilde{E}}_{lml^\prime m^\prime}=M^{EE}_{lml^\prime m^\prime}-M^{TE}_{l^{\prime\prime}m^{\prime\prime}lm}(M^{TT}_{l^{\prime\prime}m^{\prime\prime}l^{\prime\prime\prime}m^{\prime\prime\prime}})^{-1}M^{TE}_{l^{\prime\prime\prime}m^{\prime\prime\prime}l^\prime m^\prime}\,.
\end{equation}

Further, defining
\begin{equation}
\tilde{Q}(\hat{\mathbf{n}})=\frac{1}{2}\sum_{lm}\left[\tilde{E}_{lm}({}_{+2}Y_{lm}(\hat{\mathbf{n}})+{}_{-2}Y_{lm}(\hat{\mathbf{n}}))+iB_{lm}({}_{+2}Y_{lm}(\hat{\mathbf{n}})-{}_{-2}Y_{lm}(\hat{\mathbf{n}}))\right]\,,
\end{equation}
\begin{equation}
\tilde{U}(\hat{\mathbf{n}})=\frac{i}{2}\sum_{lm}\left[\tilde{E}_{lm}({}_{+2}Y_{lm}(\hat{\mathbf{n}})-{}_{-2}Y_{lm}(\hat{\mathbf{n}}))+iB_{lm}({}_{+2}Y_{lm}(\hat{\mathbf{n}})+{}_{-2}Y_{lm}(\hat{\mathbf{n}}))\right]\,.
\end{equation}
we get $ \vev{\tilde{Q}(\hat{\mathbf{n}}_i)T(\hat{\mathbf{n}}_j)}=0\,,\vev{\tilde{U}(\hat{\mathbf{n}}_i)T(\hat{\mathbf{n}}_j)}=0\,.$ The likelihood function can then be decomposed into temperature and polarization parts
\begin{equation}
\mathcal{L}(\mathbf{m}|S)\rd\mathbf{m}=\frac{\exp\left[-\frac{1}{2}\tilde{\mathbf{m}}^t(\tilde{S}_P+N_P)^{-1}\tilde{\mathbf{m}}\right]}{(2\pi)^{n_p}|\tilde{S}_P+N_P|^{1/2}}\rd\tilde{\mathbf{m}}\,\frac{\exp\left[-\frac{1}{2}\mathbf{T}^tS_T^{-1}\mathbf{T}\right]}{(2\pi)^{n_p/2}|S_T|^{1/2}}\rd\mathbf{T}
\end{equation}
where $\tilde{\mathbf{m}}=(\tilde{\mathbf{Q}},\tilde{\mathbf{U}})$. The noise in the temperature has been ignored in the above equation. The noise matrix for the new $\tilde{Q}$ and $\tilde{U}$ variables is the same as the original $Q$ and $U$ noise matrix. The new signal covariance matrix can be calculated as follows (see Appendix D of \cite{Page:2006hz})
\begin{eqnarray}
&&\vev{\tilde{Q}(\hat{\mathbf{n}}_i)\tilde{Q}(\hat{\mathbf{n}}_j)}=\frac{1}{4}\sum_{lml^\prime m^\prime}\left[M^{\tilde{E}\tilde{E}}_{lml^\prime m^\prime}({}_{+2}Y_{lm}(\hat{\mathbf{n}}_i)+{}_{-2}Y_{lm}(\hat{\mathbf{n}}_i))\right. \nonumber \\
&&\left.({}_{+2}Y_{l^\prime m^\prime}^*(\hat{\mathbf{n}}_j)+{}_{-2}Y_{l^\prime m^\prime}^*(\hat{\mathbf{n}}_j))-M^{BB}_{lml^\prime m^\prime}({}_{+2}Y_{lm}(\hat{\mathbf{n}}_i)-{}_{-2}Y_{lm}(\hat{\mathbf{n}}_i))\right. \nonumber \\
&&\left.({}_{+2}Y_{l^\prime m^\prime}^*(\hat{\mathbf{n}}_j)-{}_{-2}Y_{l^\prime m^\prime}^*(\hat{\mathbf{n}}_j))\right]\,,
\end{eqnarray}
\begin{eqnarray}
&&\vev{\tilde{Q}(\hat{\mathbf{n}}_i)\tilde{U}(\hat{\mathbf{n}}_j)}=\frac{i}{4}\sum_{lml^\prime m^\prime}\left[M^{\tilde{E}\tilde{E}}_{lml^\prime m^\prime}({}_{+2}Y_{lm}(\hat{\mathbf{n}}_i)+{}_{-2}Y_{lm}(\hat{\mathbf{n}}_i))\right. \nonumber \\
&&\left.({}_{+2}Y_{l^\prime m^\prime}^*(\hat{\mathbf{n}}_j)-{}_{-2}Y_{l^\prime m^\prime}^*(\hat{\mathbf{n}}_j))-M^{BB}_{lml^\prime m^\prime}({}_{+2}Y_{lm}(\hat{\mathbf{n}}_i)-{}_{-2}Y_{lm}(\hat{\mathbf{n}}_i))\right. \nonumber \\
&&\left.({}_{+2}Y_{l^\prime m^\prime}^*(\hat{\mathbf{n}}_j)+{}_{-2}Y_{l^\prime m^\prime}^*(\hat{\mathbf{n}}_j))\right]\,,
\end{eqnarray}
\begin{eqnarray}
&&\vev{\tilde{U}(\hat{\mathbf{n}}_i)\tilde{U}(\hat{\mathbf{n}}_j)}=-\frac{1}{4}\sum_{lml^\prime m^\prime}\left[M^{\tilde{E}\tilde{E}}_{lml^\prime m^\prime}({}_{+2}Y_{lm}(\hat{\mathbf{n}}_i)-{}_{-2}Y_{lm}(\hat{\mathbf{n}}_i))\right. \nonumber \\
&&\left.({}_{+2}Y_{l^\prime m^\prime}^*(\hat{\mathbf{n}}_j)-{}_{-2}Y_{l^\prime m^\prime}^*(\hat{\mathbf{n}}_j))-M^{BB}_{lml^\prime m^\prime}({}_{+2}Y_{lm}(\hat{\mathbf{n}}_i)+{}_{-2}Y_{lm}(\hat{\mathbf{n}}_i))\right. \nonumber \\
&&\left.({}_{+2}Y_{l^\prime m^\prime}^*(\hat{\mathbf{n}}_j)+{}_{-2}Y_{l^\prime m^\prime}^*(\hat{\mathbf{n}}_j))\right]
\end{eqnarray}
where $M^{\tilde{E}\tilde{E}}_{lml^\prime m^\prime}$ is given above by (\ref{new_EE}). The $M$ matrices are defined in \ref{cmb_maps_ap}.

The polarization part of the likelihood can be written as follows \cite{Page:2006hz}
\begin{equation}\label{like_pol_final}
\mathcal{L}(\tilde{\mathbf{m}}|\tilde{S}_P)\rd\tilde{\mathbf{m}}=\frac{\exp\left[-\frac{1}{2}(N_P^{-1}\tilde{\mathbf{m}})^t(N_P^{-1}\tilde{S}_PN_P^{-1}+N_P^{-1})^{-1}(N_P^{-1}\tilde{\mathbf{m}})\right]}{(2\pi)^{n_p}|N_P^{-1}\tilde{S}_PN_P^{-1}+N_P^{-1}|^{1/2}}|N_P|^{-1}\rd\tilde{\mathbf{m}}
\end{equation}
which is numerically more tractable since it contains only $N_P^{-1}$.

Since the calculation is done in pixel space, the formalism described above remains unchanged for a masked sky. We simply ignore the masked pixels.

More details on low-$l$ likelihood calculation can be found in the Appendix D of \cite{Page:2006hz}.

\subsubsection{Temperature Likelihood}

The temperature part of the likelihood is implemented in the class \func{Likelihood} in \file{likelihood.hpp}. To aid the numerical regularization of matrix inversion one needs to add small noise to the signal map \cite{Hinshaw:2006gs,2007ApJ...656..641E}. The class \func{Likelihood} therefore takes as an input the noise covariance matrix in addition to the signal covariance matrix. The noise part is usually added by hand by simulating white noise. For this reason the function \func{calculate} allows for input a noise map in addition to the temperature map. If the noise is already included in the temperature map, the noise map given to the function \func{calculate} should have $0$ in all the pixels.

The covariance matrices passed to the class \func{Likelihood} are in pixel space. The pixel space covariance matrices (\ref{cov_mat_pix}) are implemented in the class \func{CMatrix} in \file{c\_matrix.hpp}. The pixel space covariance matrix can be generated from the power spectrum $C_l^{TT}$ using the function \func{CMatrixGenerator::clToCMatrix} in \file{c\_matrix\_generator.hpp}. The library also provides functionality for the more general case of non-diagonal covariance matrices in harmonic space, which arise if the isotropy of space is broken. The covariance matrix in harmonic space (\ref{cov_mat_harm}) is described by the class \func{WholeMatrix} in \file{whole\_matrix.hpp}. The function \func{CMatrixGenerator::wholeMatrixToCMatrix} in \file{c\_matrix\_generator.hpp} can be used to convert a general harmonic space covariance matrix into pixel space, which can also perform a rotation by given Euler angles. This is a very useful feature for analyzing anisotropic models of the universe.

The class \func{Likelihood} takes a fiducial matrix as an input as well, which is simply added to the signal matrix. 
The fiducial matrix is commonly used to include high variance monopole and dipole terms, and higher $l$ terms not included in the signal matrix.\footnote{The signal covariance matrix might not contain all of the $l$ values that the low resolution map being analyzed is sensitive to. 
  For example, if one is analyzing a map with $N_{side}=16$, the map will be sensitive to $l\sim 2N_{side}=32$. So if the signal covariance matrix contains $l$ terms up to $l_\mathrm{max}=30$, for example, then one needs to include higher $l$ terms into the fiducial matrix. 
  The current implementation of \code{Cosmo++} will include terms up to $l = 4N_{side}$ in the fiducial matrix (starting from $l_\mathrm{max} + 1$ for the signal covariance matrix), following the WMAP likelihood code \cite{Jarosik:2010tx,Larson:2010ji,Komatsu:2010in,2013ApJS..208...19H}.} 
  Such a fiducial matrix can be generated using the function \func{CMatrixGenerator::getFiducialMatrix} in \file{c\_matrix\_generator.hpp}.

A uniform white noise matrix can be generated by \func{CMatrixGenerator::gene\hyp{}rateNoiseMatrix} in \file{c\_matrix\_generator.hpp}. The white noise map itself can be simulated by \func{Simulate::simulateWhiteNoise} in \file{simulate.hpp}.

The temperature likelihood calculation allows for foreground marginalization for a given template $\mathbf{T_f}$. A new parameter $\xi$ is introduced into the temperature likelihood function as follows
\begin{equation}\label{likelihood_fore}
\mathcal{L}_T(\mathbf{T}|C_T, \xi)=\frac{1}{(2\pi)^{n_p/2}|C_T|^{1/2}}\exp\left(-\frac{1}{2}(\mathbf{T}-\xi\mathbf{T_f})^t C_T^{-1}(\mathbf{T}-\xi\mathbf{T_f})\right)
\end{equation}
where $C_T$ denotes the total temperature covariance matrix. $\xi$ is then marginalized over 
\begin{equation}
\mathcal{L}_T(\mathbf{T}|C_T)=\int\,d\xi\,\mathcal{L}_T(\mathbf{T}|C_T, \xi)\,.
\end{equation}
The integration can be done analytically, resulting in
\begin{eqnarray}\label{likelihood_marginalized}
\mathcal{L}_T(\mathbf{T}|C_T)&=&\frac{1}{(2\pi)^{n_p/2}|C_T|^{1/2}}\sqrt{\frac{2\pi}{\mathbf{T_f}^tC_T^{-1}\mathbf{T_f}}}\nonumber \\
&&\exp\left(-\frac{1}{2}\left(\mathbf{T}^tC_T^{-1}\mathbf{T}-\frac{(\mathbf{T}^tC_T^{-1}\mathbf{T_f})^2}{\mathbf{T_f}^tC_T^{-1}\mathbf{T_f}}\right)\right)\,.
\end{eqnarray}
The user has the option of passing a foreground template to the constructor of \func{Likelihood} if she chooses to. Then (\ref{likelihood_marginalized}) will be used to calculate the temperature likelihood.

\func{Likelihood} has the option of calculating the likelihood for many maps at the same time. Since the most time consuming part of the calculation is the inversion of the covariance matrix, a lot of computing time can be saved by initializing \func{Likelihood} once followed by the likelihood calculation for many maps, instead of constructing one instance of \func{Likelihood} for each map.

\subsubsection{Polarization Likelihood}

The class \func{LikelihoodPolarization} in \file{likelihood.hpp} is used to calculate the polarization part of the likelihood, according to (\ref{like_pol_final}). The current version does not support non-zero $BB$ covariance matrices. The constructor takes as an input the pixel space signal covariance matrix $\tilde{S}_P$, the inverse noise matrix $N_P^{-1}$, as well as $M^{TE}_{l^\prime m^\prime lm}(M^{TT}_{l^\prime m^\prime l^{\prime\prime}m^{\prime\prime}})^{-1}$ and $T_{lm}$ in order to be able to calculate (\ref{e_redef}). The function \func{combineWholeMatrices} in \func{LikelihoodPolarization} can be used to construct $M^{TE}_{l^\prime m^\prime lm}(M^{TT}_{l^\prime m^\prime l^{\prime\prime}m^{\prime\prime}})^{-1}$ and $M^{\tilde{E}\tilde{E}}_{lml^\prime m^\prime}$ from $M^{TT}$, $M^{TE}$ and $M^{EE}$. $M^{\tilde{E}\tilde{E}}_{lml^\prime m^\prime}$ can then be converted into $\tilde{S}_P$ using the funciton \func{CMatrixGenerator::polarizationEEWholeMatrixToCMatrix} in \file{c\_matrix\_generator.hpp}.

As for the temperature case, non-diagonal covariance matrices in harmonic space are supported. Also, likelihood calculation for many maps at once is supported.

\subsection{High-$l$ Likelihood}

For high multipoles, the distribution of the $C_l$'s can be very well approximated by a Gaussian distribution, allowing for a much faster likelihood calculation. We follow the approach described in \cite{Hinshaw:2003hi} for the implementation of high-$l$ likelihood calculation. The current version includes high-$l$ likelihood calculation for temperature maps only.

Before calculating the likelihood function, one needs to estimate the power spectrum $\hat{C}_l^{TT}$ for the data. A few different approaches for this calculation have been described in the literature, including the MASTER algorithm \cite{Hivon:2001eh}, quadratic maximum likelihood (QML) estimators \cite{Tegmark:1996bm}, and the XFaster algorithm \cite{Rocha:2009wo}. The current version of \code{Cosmo++} includes an implementation of the MASTER algorithm \cite{Hivon:2001eh} in the class \func{Master} in \file{master.hpp}.

The first step in the power spectrum calculation is the mask coupling kernel $K_{l_1l_2}$ using the function \func{calculateCouplingKernel} in the class \func{Master}. The mask coupling kernel is defined by \cite{Hivon:2001eh}
\begin{equation}\label{mask_ck}
K_{l_1l_2}=\frac{2l_2+1}{4\pi}\sum_{l_3}(2l_3+1)M_{l_3}\left(\begin{array}{ccc}
l_1 & l_2 & l_3 \\
0 & 0 & 0 \end{array}\right)
\end{equation}
where $\left(\begin{array}{ccc}
l_1 & l_2 & l_3 \\
0 & 0 & 0 \end{array}\right)$ denotes the Wigner $3-j$ symbol,
\begin{equation}
M_{l}=\frac{1}{2l+1}\sum_{m=-l}^l|m_{lm}|^2\,,
\end{equation}
and $m_{lm}$ is the mask transformed into harmonic space. The mask coupling kernel relates the ensemble average of the masked power spectrum $\tilde{C}_l$ to the unmasked one $C_l$
\begin{equation}
\langle\tilde{C}_{l_1}\rangle=\sum_{l_2}K_{l_1l_2}\vev{C_{l_2}}\,.
\end{equation}

The calculation of the mask coupling kernel is the slowest part of the whole power spectrum calculation, therefore the user is allowed to save the result in a file and use it many times for different maps. Once this is done, the result can be passed to the constructor of \func{Master}, then the data power spectrum can be calculated for a given map using the function \func{calculate}.

The high-$l$ likelihood itself can be calculated using the class \func{LikelihoodHigh} in \file{likelihood.hpp}. This class takes as an input the output of \func{Master}, as well as the noise power spectrum $N_l$. The likelihood for a given theoretical power spectrum $C_l$ is calculated by first calculating the cut-sky Fisher matrix \cite{Hinshaw:2003hi}
\begin{equation}
  F_{l_1l_2}=\frac{(2l_1+1)}{2(C_{l_1}+N_{l_1})(C_{l_2}+N_{l_2})}\tilde{F}_{l_1l_2}
\end{equation}
where
\begin{equation}
  \tilde{F}^S_{l_1l_2}=K_{l_1l_2}
\end{equation}
in the signal dominated limit and
\begin{equation}
  \tilde{F}^N_{l_1l_2}=\frac{1}{\bar{w}^2}K^\prime_{l_1l_2}
\end{equation}
in the noise dominated limit. Here
\begin{equation}\label{noise_ck}
K^\prime_{l_1l_2}=\frac{2l_2+1}{4\pi}\sum_{l_3}(2l_3+1)W_{l_3}\left(\begin{array}{ccc}
l_1 & l_2 & l_3 \\
0 & 0 & 0 \end{array}\right)\,,
\end{equation}
\begin{equation}
W_{l}=\frac{1}{2l+1}\sum_{m=-l}^l|w_{lm}|^2\,,
\end{equation}
and $w_{lm}$ are the pixel weights transformed into harmonic space. The pixel weights are determined from the inverse noise matrix
\begin{equation}
  (N^{-1})_{ij}=w_i\delta_{ij}\,.
\end{equation}
and further multiplied with the mask (i.e. the pixel weights are set to $0$ in the masked pixels). We are assuming that the pixel noises are uncorrelated. $\bar{w}$ denotes the average pixel weight.

The coupling kernel $K^\prime_{l_1l_2}$ can be calculated using the \func{Master} class by passing the masked pixel weight map to it instead of a mask. If the likelihood calculation is done in the signal dominated regime only the calculation of $K^\prime_{l_1l_2}$ can be omitted. If this is not passed to the \func{LikelihoodHigh} class it will automatically do the calculation assuming signal domination.

The Fisher matrix in the intermediate regime is determined by interpolating between the signal dominated and noise dominated regimes \cite{Hinshaw:2003hi}
\begin{equation}
  \tilde{F}_{l_1l_2}=\frac{\left(C_{l_1}\sqrt{\tilde{F}^S_{l_1l_2}}+N_{l_1}\sqrt{\tilde{F}^N_{l_1l_2}}\right)\left(C_{l_2}\sqrt{\tilde{F}^S_{l_1l_2}}+N_{l_2}\sqrt{\tilde{F}^N_{l_1l_2}}\right)}{(C_{l_1}+N_{l_1})(C_{l_2}+N_{l_2})}
\end{equation}

After calculating the Fisher matrix the likelihood function can be calculated by
\begin{equation}
-2\ln\mathcal{L}(\hat{C}_l|C_l)=\sum_{l_1l_2}(C_{l_1}-\hat{C}_{l_1})F_{l_1l_2}(C_{l_2}-\hat{C}_{l_2})\,.
\end{equation}

The above expression assumes that $\ln\mathcal{L}=0$ for the best fit case, i.e. the likelihood is calculated up to a fixed constant factor. Since for any practical purposes only the likelihood ratios matter (or the difference in $\ln\mathcal{L}$), the constant factor can be safely ignored.


\subsection{CMB Gibbs Sampler}

Gibbs sampling has been proposed as an alternative approach to CMB power spectrum estimation and likelihood calculation \cite{Eriksen:2004kn,Groeneboom:2009ti,Eriksen:2007ht}. \code{Cosmo++} includes an implementation of the CMB Gibbs sampler in the class \func{CMBGibbsSampler} in \file{cmb\_gibbs.hpp}. The implementation follows almost exactly the algorithm described in \cite{Eriksen:2004kn}, so we will not go over the details here again.

The Gibbs sampler is usually used for likelihood calculation for low $l$ values, but it can be used for $l$ values higher than the pixel space likelihood calculation allows for. In pixel space, one can do the calculation up to $l=30$, using reduced resolution maps with HEALPix $N_{side}=16$, but for higher $l$ values one needs higher resolution maps and this significantly increases the computational costs. The main reason is that one needs to invert a matrix with size equal to the number of pixels. The Gibbs Sampler, on the other hand, does not need to obtain the inverted matrix, it only needs to solve a system of linear equations, and this can be done using the preconditioned conjugate gradient algorithm \cite{Shewchuk:1994uc}. Preconditioners have been proposed in the literature which allow for fast convergence; we use the preconditioner described in \cite{Eriksen:2004kn}, eq. (28). This allows for fast likelihood calculation up to $l=50$ and even higher values. A cut sky is handled by simply setting the inverse noise matrix elements to $0$ for masked pixels. This approach is called the Commander implementation \cite{Eriksen:2004kn}, and is used in the Planck likelihood code \cite{Collaboration:2013vc}, in particular.

\func{CMBGibbsSampler} can be used to first construct the Gibbs chain and save it in a file, after which the chain can be used for very fast likelihood calculation. The Blackwell-Rao estimator \cite{Rudjord:2008dr} is used for likelihood calculation.

\section{Parameter Space Sampling}\label{sampling_sec}

Bayesian methods have become an essential part of cosmological parameter estimation in the past two decades (for a comparison of different methods see, e.g. \cite{Allison:2013tq}). \code{Cosmo++} provides a general interface for parameter space sampling, which can be used for cosmological parameter estimation, in particular.

The abstract class \func{Math::LikelihoodFunction} in \file{likelihood\_function.hpp} provides a simple interface for likelihood calculation from a given number of parameters. The user needs to inherit a likelihood class from \func{Math::LikelihoodFunc\hyp{}tion} and implement their own likelihood calculation in the function \func{calculate}. For example, the likelihood calculation tools described in Section \ref{likelihood_sec} can be used in combination to implement a likelihood function for cosmological parameters. Section \ref{examples_sec} includes some specific examples.

\code{Cosmo++} includes two parameter space sampling tools, a basic Metropolis-Hastings sampler \cite{mcmc}, and a MultiNest sampler \cite{Feroz:2008eb,Feroz:2013uq,Feroz:2007fi}. The Metropolis-Hastings sampler has been implemented in \code{Cosmo++} itself, while the MultiNest sampler uses the publicly available \code{MultiNest} code for implementation.
 The user has the option of not including the \code{MultiNest} code, then \code{Cosmo++} will be compiled without the MultiNest sampler.

The Metropolis-Hastings sampler is implemented in the class \func{Math::Metropo\hyp{}lisHastings} in \file{mcmc.hpp}, while the MultiNest scanner interface is implemented in the class \func{MnScanner} in \file{mn\_scanner.hpp}. The two classes have almost identical interfaces, allowing the user to easily switch between them. The only differences in the interfaces are due to some parameters that are specific for each one of them. Both of the constructors take an instance of \func{Math::LikelihoodFunction} to set the likelihood function, then the parameters need to be set, including their name, range, and the prior function. After this the function \func{run} can be called to do the actual scan. The results are written in text files.

Both of the samplers allow the user to set a uniform or a Gaussian prior on the parameters. \func{Math::MetropolisHastings} allows the user to set an external prior function for all of the parameters in case a more general function is needed to be used or the priors of the different parameters are not independent. The \func{MultiNest} sampler also allows for general priors on the parameters, however each parameter prior needs to be independent of the others. By default, \func{Math::MetropolisHastings} uses Gaussian proposal distributions, with widths that can be chosen by the user, or are set to $1/100$-th of the parameter range by default. The user also has the option of setting their own proposal distribution. The parameters are varied in blocks. Each block contains one parameter by default, but the user has the option of setting their own blocks. Both of the scanners have the option of resuming from the point they stopped.

Both of the samplers support an MPI implementation. The implementation of the MultiNest sampler is described in \cite{Feroz:2008eb,Feroz:2013uq,Feroz:2007fi}. The MPI implementation of the Metropolis-Hastings sampler generates one independent chain per MPI instance. The starting values of the parameters are distributed around the given value specified by the user with a given width. The MPI processes send updates to the ``master'' after every $100$ iterations. The stopping time is determined from the updates of all of the processes.

There are two stopping criteria implemented for \func{Math::MetropolisHastings}. The first criterion is called ``Accuracy'' for which the stopping time is determined from the given accuracies of the parameters. Namely, the run stops when the standard deviation of the mean of all of the parameters becomes less than their accuracies. The standard deviation of the mean is calculated taking into account the autocorrelation between the elements of the chain. The second criterion is the Gelman-Rubin criterion \cite{Gelman:1992ts}, which requires more than one chain. The Gelman-Rubin criterion is the recommended stopping criterion. However, if the user selects the Gelman-Rubin criterion and runs the engine with only one MPI process, the stopping criterion will automatically be switched to ``Accuracy'', since the Gelman-Rubin criterion needs more than one chain. A maximum length of the chain must also be given to the \func{run} function and the run will always stop when that maximum length is reached (in case of multiple chains, the run will stop when the ``master'' process reaches the maximum chain length), even if the requested accuracies are not reached. This allows the user to use other stopping criteria which can estimate the chain length required for convergence \cite {mcmc}.

The functionality of the MultiNest scanner is described in detail in \cite{Feroz:2008eb,Feroz:2013uq,Feroz:2007fi}. Our interface \func{MnScanner} gives two options for determining the stopping time. The user can specify whether or not they would like accurate Bayesian evidence calculation, then the code selects the recommended MultiNest parameters for each case \cite{Feroz:2008eb}.

The resulting chain files have the standard format used in cosmology codes, such as \code{CosmoMC} \cite{Lewis:2002hg} and \code{MultiNest} \cite{Feroz:2008eb,Feroz:2013uq,Feroz:2007fi}. The user can analyze the chains using the \func{MarkovChain} class from the file \file{markov\_chain.hpp}. This class allows the user to read one or more chains from given files (with appropriate burnins and thinning factors), then generate marginalized one or two dimensional distributions for the parameters. These distributions are stored in the classes \func{Posterior1D} and \func{Posterior2D}, respectively. From the posterior distributions the user is able to obtain various statistical quantities, such as the mean and median values, as well as different confidence intervals. The distributions can also be output into files and plotted. The sample python scripts \file{function\_plot.py} and \file{contour\_plot.py} are provided, which can be used to plot one dimensional marginalized distributions and two dimensional contours. For an example of running a parameter space sampler and plotting the resulting distributions see Section \ref{param_space_sampling_example_sec}.

There are also publicly available standard tools that can be used to analyze the resulting chains, such as \code{getdist} included in \code{CosmoMC} and the python package \code{Pippi} \cite{Scott:2012em}.

\section{Planck and WMAP Likelihood}\label{planck_wmap_like_sec}

The Planck likelihood code \cite{Collaboration:2013vc} is publicly available and can be used as a library to be linked to other code. Since most of the current research in cosmology uses this code, we found it useful to include an interface in \code{Cosmo++}. This also serves as a useful example of implementing an instance of the abstract class \func{Math::LikelihoodFunction} described in Section \ref{sampling_sec}, which can be directly used by the parameter space samplers. The user may choose to compile \code{Cosmo++} without linking to the Planck likelihood code in which case the functionality described above will be absent.

The Planck likelihood interface is implemented in the \func{PlanckLikelihood} class in \file{planck\_like.hpp}. Our implementation allows the user to set the cosmological parameters first, then any extra parameters that are used for modelling the foreground effects, after which different Planck likelihoods can be calculated. The user can also calculate the combination of all of the likelihoods that were chosen in the constructor through the function \func{calculate}. This function is used by the parameter space samplers described in Section \ref{sampling_sec}. The \func{PlanckLikelihood} class always checks if the values of the cosmological parameters given have changed since the last call to decide whether or not to re-calculate the CMB power spectra.

\code{Cosmo++} also includes an interface for the WMAP nine-year likelihood code \cite{Jarosik:2010tx,Larson:2010ji,Komatsu:2010in,2013ApJS..208...19H} in the \func{WMAP9Likelihoood} class in \file{wmap9\_like.hpp}. The functionality and the interface are very similar to \func{PlanckLikelihood}. In particular, an instance of this class can be passed to the parameter space samplers, and the user is allowed to choose which likelihoods (temperature, polarization, low-$l$, high-$l$) to include in the calculation.

\section{Simulations}\label{sim_sec}

\code{Cosmo++} provides the functionality for simulating sky maps in harmonic space. This is implemented in the class \func{Simulate} in \file{simulate.hpp}. The maps in harmonic space can be converted to pixel space using the \func{map2alm} function in the \code{HEALPix} C++ package.

The user has the option of simulating the sky from given $C_l$ values or from the full covariance matrix $M_{lml^\prime m^\prime}$. The first case is similar to the \code{HEALPix} tool \code{synfast}, the second case is more general and allows the user to do simulations for anisotropic universes with non-diagonal covariance matrices. The non-diagonal case is implemented by switching to a basis in which the matrix becomes diagonal, then doing the simulation in the new basis, and finally switching back to the original basis. This means that the computational time and the memory requirements will increase significantly as $l$ increases.

The class \func{Simulate} also provides the functionality of simulating uniform white noise maps in pixel space. This can be done using the function \func{simulateWhiteNoise}.

\section{Other Utilities}\label{utilities_sec}

\subsection{Mask Apodization}

Masks are commonly used in cosmology to separate out the reliable data points. For example, when using CMB data, one must mask out the plane of our galaxy, as well as some bright sources that cannot be reliably subtracted out from the radiation data. Sharp edges of the mask in pixel space may introduce undesirable effects when the calculation is done in harmonic space \cite{Collaboration:2013vc}. For this reason it is sometimes necessary to apodize the mask, i.e. smooth out the edges of the mask. This functionality is included in \code{Cosmo++} through the class \func{MaskApodizer} in \file{mask\_apodizer.hpp}. The input and output masks are in \code{HEALPix} format.

Two apodization types are supported: cosine and Gaussian. For the cosine apodization the $1$ mask values near the edge are replaced by $1 - \cos(\theta\pi/2\theta_{ap})$ if $\theta\le\theta_{ap}$. Here $\theta$ denotes the angular distance from the edge, $\theta_{ap}$ is the apodization angle. In case of Gaussian apodization the $1$ values are replaced by $1 - \exp(-(3\theta)^2/2\theta_{ap}^2)$ if $\theta\le\theta_{ap}$.

The apodization routine is implemented by first finding the pixels that are on the edge, then by scanning through all of the masked points and finding the nearest distance from the edge. The first stage is linear in the number of pixels and is relatively faster. The second stage involves $O(N_\mathrm{masked}N_\mathrm{edge})$ operations, where $N_\mathrm{masked}$ is the number of masked pixels, $N_\mathrm{edge}$ is the number of pixels that lie on the edge. The computational speed of this stage depends strongly on the geometry of the mask.

\subsection{Angular Momentum Dispersion Calculation}

\code{Cosmo++} includes a tool for calculating the angular momentum dispersion $\sum_m m^2|T_{lm}(\mathbf{\hat{n}})|^2$ as a function of the direction $\mathbf{\hat{n}}$, where $T_{lm}(\mathbf{\hat{n}})$ denote the spherical harmonic coefficients in a coordinate system where the $z$-axis has $\mathbf{\hat{z}}||\mathbf{\hat{n}}$. The main use of the angular momentum dispersion calculation is to determine the alignment of different multipoles \cite{deOliveiraCosta:2003je,Aslanyan:2013we}. This calculation is implemented in the class \func{ModeDirections} in \file{mode\_directions.hpp}. The calculation for a given direction can be done using the \func{calculateAngularMomentumDispersion} function. One can also find the direction which maximizes the angular momentum dispersion with the function \func{maximizeAngularMomentumDispersion}.

\section{Mathematical Tools}\label{tools_sec}

\code{Cosmo++} includes a few additional mathematical tools that are used in the implementation of the features described above. These tools are fully documented as well and can be used independently. Most of the mathematical tools are fully implemented in \code{Cosmo++} and are independent of any external libraries. If the user needs to use the mathematical tools only then she can simply compile \code{Cosmo++} as it is, without including any external libraries.

\subsection{Interpolation}

\code{Cosmo++} includes classes that provide interpolation functionality between given points. Two interpolation methods have been implemented: linear interpolation and cubic spline. Given a number of knots, the linear interpolation algorithm constructs a continuous (but not differentiable at the knots) function passing through the knots by simply connecting them by line segments. The cubic spline algorithm, on the other hand, constructs a smooth function passing through these points. The points are connected by a piecewise cubic polynomial, and the resulting curve has continuous first and second derivatives.

The linear interpolation method has been implemented in the \func{Math::Table\hyp{}Function} class in \file{table\_function.hpp}. This class is derived from the standard C++ class \func{std::map}, which gives it the full interface of \func{std::map}. In particular, this allows the user to define the data points using the regular interface of \func{std::map}. \func{Math::TableFunction} is also a child of the abstract class \func{Math::Function}, which means that an object of this type can be passed to the rest of the library wherever a one-variable function is needed. For example, an instance of \func{Math::TableFunction} can be used directly to represent a primordial power spectrum (see Section \ref{ps_sec}). The linear interpolation method has been generalized to two and three dimensions in the classes \func{Math::TableFunction2} and \func{Math::TableFunction3}, respectively. They are both in the same header file \file{table\_function.hpp}.

The cubic spline interpolation has been implemented in the \func{Math::CubicSpline} class in \file{cubic\_spline.hpp}. This class is derived from \func{Math::RealFunction}, allowing the user to pass an instance of \func{Math::CubicSpline} to the rest of the library wherever a one-variable real function is needed.\footnote{Note that the \func{Math:RealFunction} type is the same as \func{Math::Function$<$double, double$>$}, i.e. it is a specific case of \func{Math::Function} with the variable types fixed to be real numbers.}

\subsection{Conjugate Gradient Solver}

The preconditioned conjugate gradient method \cite{Shewchuk:1994uc} for solving a system of linear equations has been implemented in \code{Cosmo++} in the class \func{Math::Conju\hyp{}gateGradient} in \file{conjugate\_gradient.hpp}. Since the method can be used for potentially very big matrices, we have implemented the matrix multiplication and the preconditioning to be performed through a template type that the user needs to implement. This gives the user complete freedom in implementing the matrix multiplication. The entire matrix does not have to be stored in the memory, the user may choose to calculate each matrix element as the multiplication is performed or apply the linear operator on the vector in any other way. The simple case when the matrix and the preconditioner are known and can be set initially is implemented in the \func{Math::BasicCGTreats} class.

\subsection{Legendre Polynomials and Spherical Harmonics}

The \func{Math::Legendre} and \func{Math::AssociatedLegendre} classes in \file{legendre.hpp} can be used to calculate the Legendre polynomials and associated Legendre polynomials, respectively. The \func{Math::SphericalHarmonics} class in \file{spherical\_harmonics.hpp} is a spherical harmonics calculator. All of these classes are very easy to use. The constructors do not take any arguments, and the \func{calculate} function in each class simply calculates the result.

\subsection{Rotations in Three Dimensions}

$O(3)$ rotations, i.e. rotations of three dimensional real vectors, have been implemented in the \func{Math::ThreeRotationMatrix} class in \file{three\_rotation.hpp}. The class \func{Math::ThreeVector} represents three dimensional vectors, which can be passed as an input to \func{Math::ThreeRotatioMatrix} to be rotated. The user can set the rotation matrix either by specifying three Euler angles $(\phi, \theta, \psi)$, or by giving an axis and an angle of rotation around that axis. The user can then read the matrix elements, or perform a rotation on a given vector. The multiplication operation has been defined for these matrices.

By default, the rotation is passive, i.e. the coordinate frame is rotated rather than the vector. The convention for Euler angles is as follows. First a counterclockwise rotation is performed around the $z$ axis by angle $\phi$, then a counterclockwise rotation around the new $x$ axis by angle $\theta$, and finally a counterclockwise rotation around the new $z$ axis by angle $\psi$.

\subsection{Wigner $3-j$ Symbols}

The main method for computing the Wigner $3-j$ symbols is by using recursion relations, and this is how most of the numerical libraries implement the calculation. Most of the time, however, one needs to calculate a lot of these symbols at once. Numerical libraries, such as the \code{GNU Scientific Library}\footnote{http://www.gnu.org/software/gsl/}, have functions for calculating each symbol by itself. When one calls these functions many times for a range of indices, a lot of the calculation is repeated many times, because to calculate the values for higher indices these functions need to calculate all of the lower index symbols again to use recursion. We have solved this problem in \code{Cosmo++} by creating an interface that allows the user to first set the maximum value for the indices, after which the symbols are sent back one by one as they are being calculated. This allows for the whole calculation to be performed only once. Compared to the case when each symbol is calculated from scratch, our approach gives a huge improvement in computational time. We use this implementation to calculate the mask coupling kernel (\ref{mask_ck}), for example, where all of the symbols are needed with indices up to a given maximum value.

The current version of \code{Cosmo++} only includes the implementation for the case when all of the $m$ indices are $0$. This is done in the template class \func{Math::Wigner3JZeroM} in \file{wigner\_3j.hpp}. The template parameter is a simple class that needs to include a \func{process} function. The symbols are then sent back to the user by calling the \func{process} function for an object given by the user.

\subsection{Random Number Generator}

\code{Cosmo++} includes three random number generators: a real number generator with uniform distribution \func{Math::UniformRealGenerator}, a Gaussian distribution generator \func{Math::GaussianGenerator}, and a Poisson distribution generator \func{Math::PoissonGenerator}. All of these classes are defined in \file{random.hpp}. The implementation uses the C++ standard library (available starting the C++11 standard). The advantage of using these classes throughout the code as compared to using the standard library functions directly is that the user can easily switch to a different implementation for these cases by simply modifying the random number generator classes defined here. This way the user will not have to go through the code and find all of the places random number generators are used.

%

\subsection{Curve Fitting}

Basic curve fitting functionality is included in \code{Cosmo++} through the class \func{Math::Fit} in \file{fit.hpp}.  The function to be fit to the data points is passed as a parametric function, and the number of the parameters is a template argument. Specifically, a child class of the class \func{Math::ParametricFunction} needs to be implemented and passed to \func{Math::Fit} as an input. As a useful case, we have implemented a polynomial parametric function (the parameters are the coefficients of the polynomial) in the class \func{Math::Polynomial} in \file{polynomial.hpp}.

The curve fitting is done by simply minimizing $\chi^2=\sum_i(y_i-f(x_i))^2$ as a function of the parameters. Here $(x_i, y_i)$ denote the data points, $f$ denotes the function to be fit. The minimization is done using the publicly available C++ package \code{Minuit}\footnote{http://seal.web.cern.ch/seal/work-packages/mathlibs/minuit}.

%

\section{Examples}\label{examples_sec}

We have implemented a few examples that can help the user better understand the functionality of the library. Some of these examples may be useful on their own. All of the examples are thoroughly commented.

\begin{figure}[t]
\centering
\includegraphics[width=6cm]{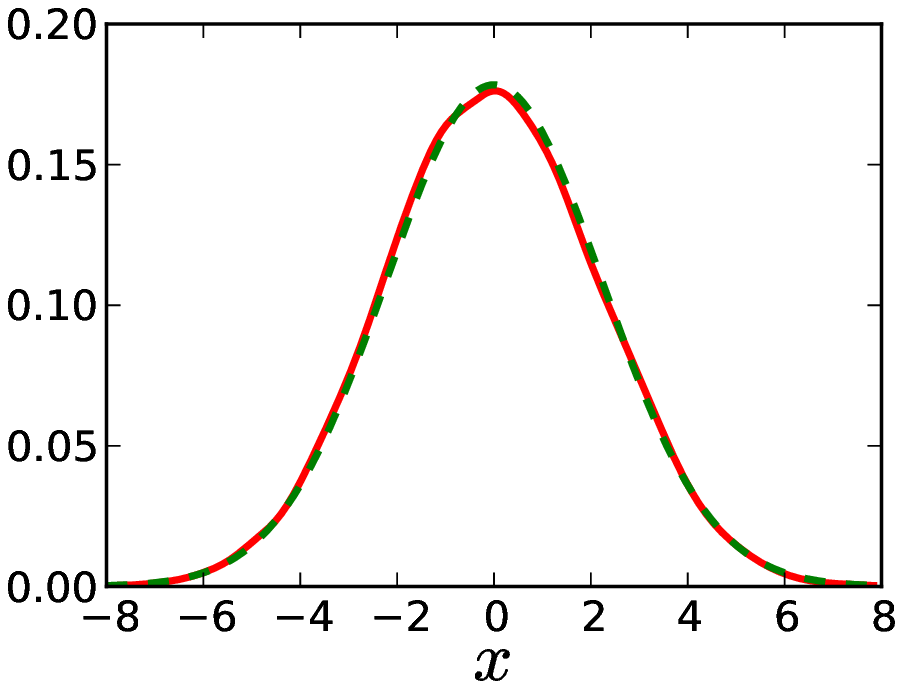}
\includegraphics[width=6cm]{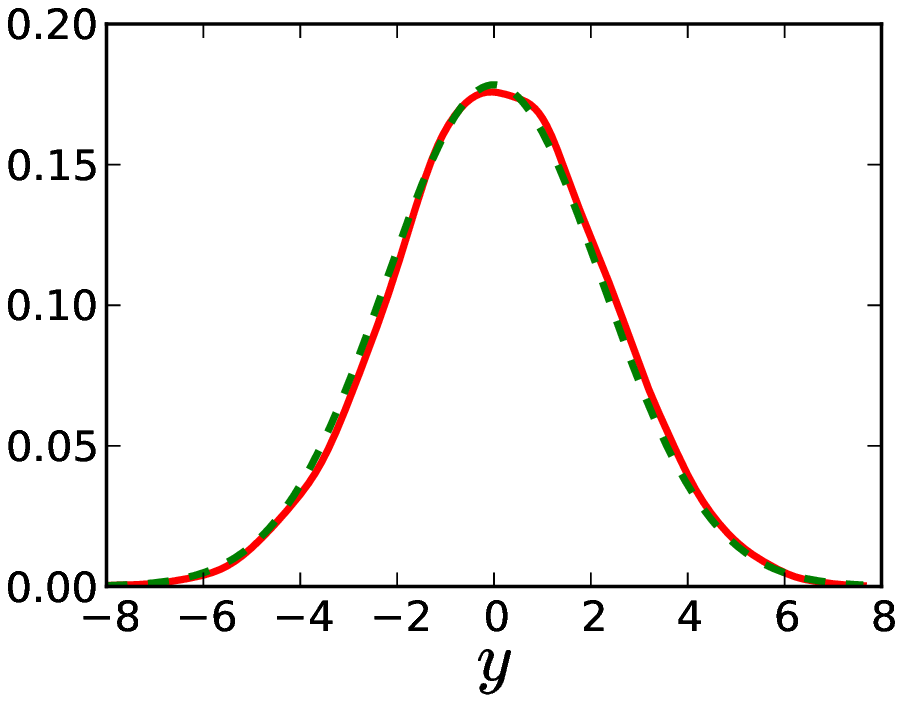}
\includegraphics[width=6cm]{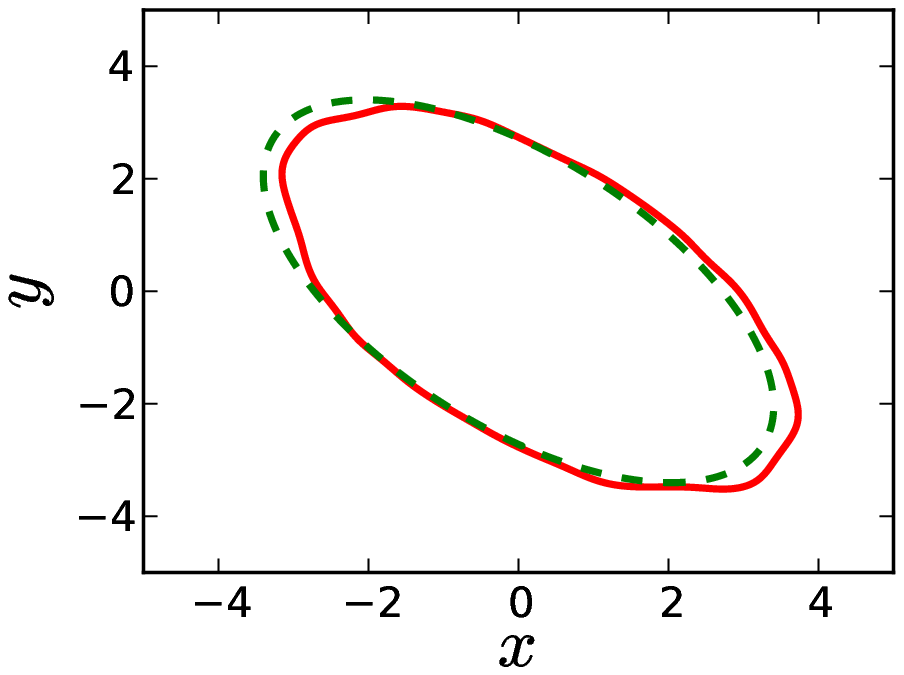}
\caption{\label{mh_example_fig} The marginalized parameter distributions for the example likelihood (\ref{example_likelihood}) from the Metropolis-Hastings sampler (top) and the $1\,\sigma$ contour plot (bottom). The red solid lines show the results, the greed dashed lines are the theoretical expectations from (\ref{example_likelihood}).
}
\end{figure}

\subsection{Parameter Space Sampling}\label{param_space_sampling_example_sec}

Our first example \file{example\_metropolis\_hastings.cpp} simply demonstrates how to implement a two parameter likelihood function and pass it into the Metropolis-Hastings sampler. After running the sampler, the resulting chain is analyzed and the marginalized one dimensional distributions of the parameters, as well as the joint two-dimensional distribution, are written into text files. These distributions are then plotted with the provided sample python scripts \file{mh\_example\_x\_plot.py}, \file{mh\_example\_y\_plot.py}, and \file{mh\_example\_contour.py}.

In our example we use a bivariate normal distribution:
\begin{equation}\label{example_likelihood}
  \mathcal{L} = \frac{1}{2\pi\sigma_1\sigma_2}\exp\left(-\frac{((x + y) / 2)^2}{2\sigma_1^2}-\frac{((x - y) / 2)^2}{2\sigma_2^2}\right)
\end{equation}
where $\sigma_1 = 1$, $\sigma_2 = 2$. The marginalized parameter distributions and the $1\,\sigma$ contour plot are shown in Figure \ref{mh_example_fig}. The results are compared to the expected curves obtained from (\ref{example_likelihood}) which are shown as dashed lines. As can be seen in the plots, the results agree very well with the expected distributions.

\subsection{CMB Power Spectra Calculation}

The example \file{example\_cl.cpp} includes a simple calculation of CMB power spectra. The values of the standard cosmological parameters are defined, then an instance of the \func{CMB} class is created with these parameters, the power spectra are calculated, and the results are written in text files. Upon compilation, this file is turned into the executable \func{example\_cl}.

\subsection{Planck Likelihood}

A simple example of using the Planck likelihood module is implemented in \file{example\_planck.cpp}. An instance of the cosmological parameters' class is created with specific values of the parameters and passed to the Planck likelihood class. After this different Planck likelihoods are calculated and printed on the screen.

\subsection{Planck Likelihood and MultiNest}

As a second example of parameter space sampling we have implemented a MultiNest sampler for the Planck likelihood code in \file{example\_mn\_planck.cpp}. After compiling \code{Cosmo++} turns this into the executable \func{example\_mn\_planck} which can be used directly to calculate the posterior distributions and the confidence intervals of the standard cosmological parameters from Planck. The example contains only about $50$ lines of code, and is very straightforward to change and generalize.

\section{Tests}\label{tests_sec}

The \code{Cosmo++} library includes a test framework class \func{TestFramework} that can be used to easily create unit tests. Rigorous tests for all of the modules of \code{Cosmo++} itself have been created using this module. After compiling the library, the executable \func{test} can be called to get a list of all of the unit test, as well as run them. The tests are divided into two categories - fast and slow. The user can choose to run a test by name, all of the tests in one category, as well as all of the tests together. The fast tests take only a few seconds while each of the slow tests can take up to a few days. For this reason it is recommended to run each slow test by itself and possibly use many MPI nodes.

In this section we describe the test framework and some of the most important tests of \code{Cosmo++}.

\subsection{The Test Framework}

Each test can have a number of subtests. A subtest needs to be simple enough so that the result is one real number that should be compared to the expected value. In order to create a unit test the user needs to inherit their test class from \func{TestFramework} and implement the purely virtual functions \func{name}, \func{numberOfSubtests}, and \func{runSubTest}. The names are self-explanatory. The main functionality will be implemented in the \func{runSubTest} function. Each subtest needs to return two real numbers - an actual result and an expected result, as well as the name of the subtest.

The test can be run by calling the \func{run} function. The precision with which the actual results are compared to the expected results can be chosen in the constructor.

\subsection{Mathematical Tools}

We have separately tested all of the mathematical tools described in Section \ref{tools_sec}. Unless specified otherwise, the precision of all of the tests of the mathematical tools is $10^{-5}$.

The class \func{TestThreeRotations} tests the three dimensional rotation matrices. The test applies a few sample rotations to given vectors and compares the resulting vector coordinates to the expected values.

The class \func{TestLegendre} tests the Legendre and associated Legendre polynomial calculators and the class \func{TestSphericalHarmonics} tests the spherical harmonics calculator. The tests calculate some sample values and compare to the results obtained from the \code{Mathematica} software\footnote{http://www.wolfram.com/mathematica/}. The maximum $l$-value tested in all the cases is $10000$.

The class \func{TestWigner3J} tests the Wigner $3-j$ symbol calculator. A few sample cases are calculated and compared to the results from \code{Mathematica}. The maximum $l$-value tested is $1000$.

The classes \func{TestTableFunction} and \func{TestCubicSpline} test the linear and cubic spline interpolations, respectively. They simply interpolate between a few fixed points and calculate the interpolation at some points in between the interpolation points.

The class \func{TestFit} tests the curve fitting routine. It first constructs a third degree polynomial with given coefficients, then selects four different points on it, then fits a third degree polynomial to these points and compares the coefficients obtained from the fit to the original polynomial coefficients. These should agree exactly since it is always possible to exactly fit a third degree polynomial to four different points.

The class \func{TestConjugateGradient} tests the conjugate gradient solver. It constructs symmetric positive definite matrices of different sizes, multplies them with given vectors, and feeds the resulting vector, as well as the matrix to the conjugate gradient solver. The resulting solution is compared to the original vector. The maximum matrix size tested is $2500\times2500$. For this case it takes only a few seconds on a single core to reach the precision of $10^{-7}$.

\subsection{Parameter Space Samplers}

We have created fast unit tests for both the Metropolis-Hastings sampler and the MultiNest sampler in the classes \func{TestMCMCFast} and \func{TestMultinestFast}, respectively. Both of them create a two variable Gaussian likelihood function, then do the scan with the chosen parameter space sampler, then analyze the resulting chains to get the marginalized posterior distributions for both of the variables. The resulting confidence ranges for the variables are then compared to the expected ones.

\subsection{Mask Apodization}\label{mask_ap_test_sec}

The class \func{TestMaskApodizer} tests the mask apodization routine. It constructs a simple galactic mask, apodizes it, then checks the value of a selected pixel in the apodization region. The test is performed for both cosine and Gaussian apodization schemes. The test mask being apodized has \code{HEALPix} $N_{side}=2048$, and the apodization angle is $\pi / 10$. With these settings the apodized pixel values agree with the expected values with a precision of $10^{-3}$.

\begin{figure}
\centering
\includegraphics[width=6cm]{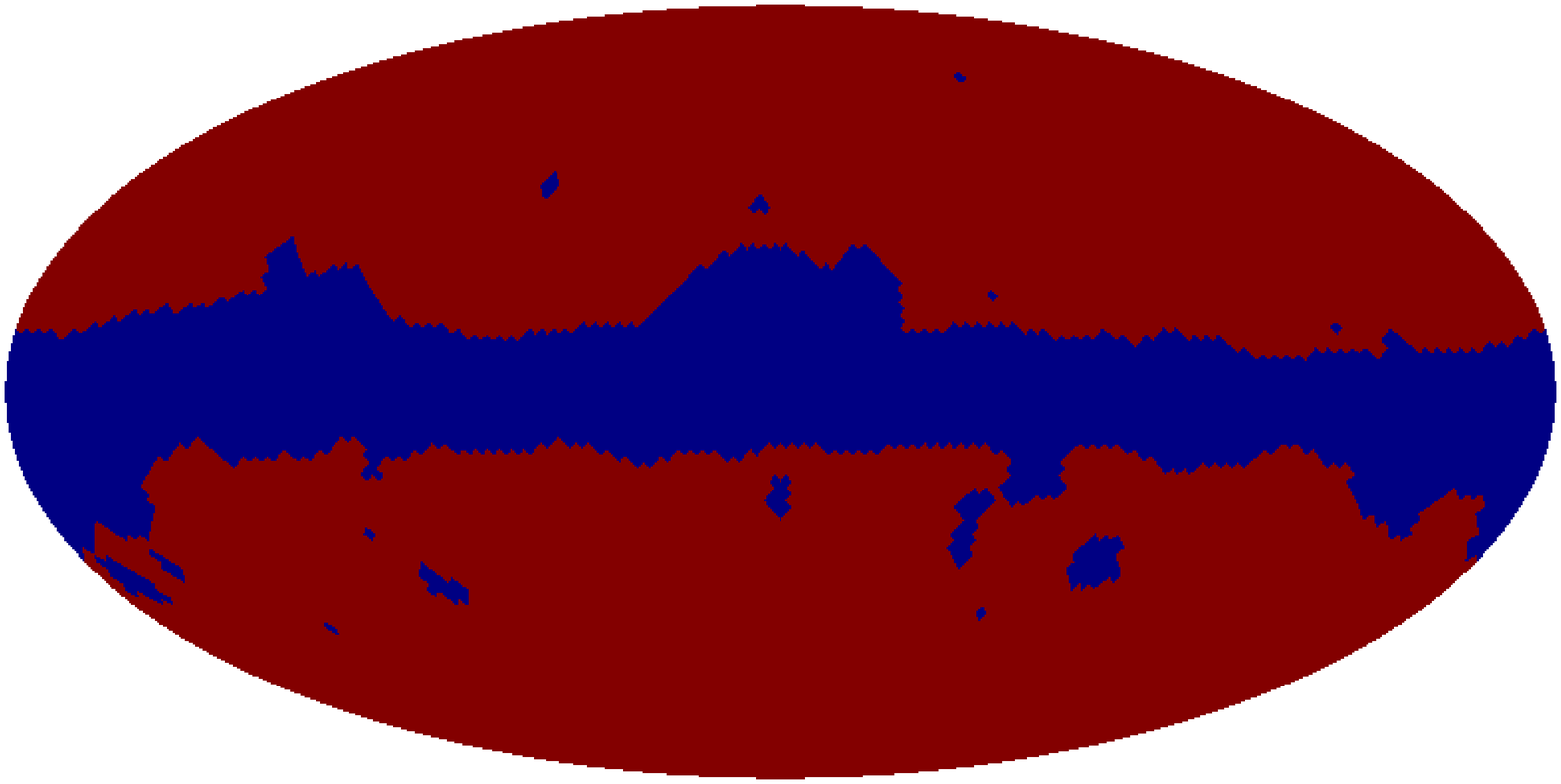}
\includegraphics[width=6cm]{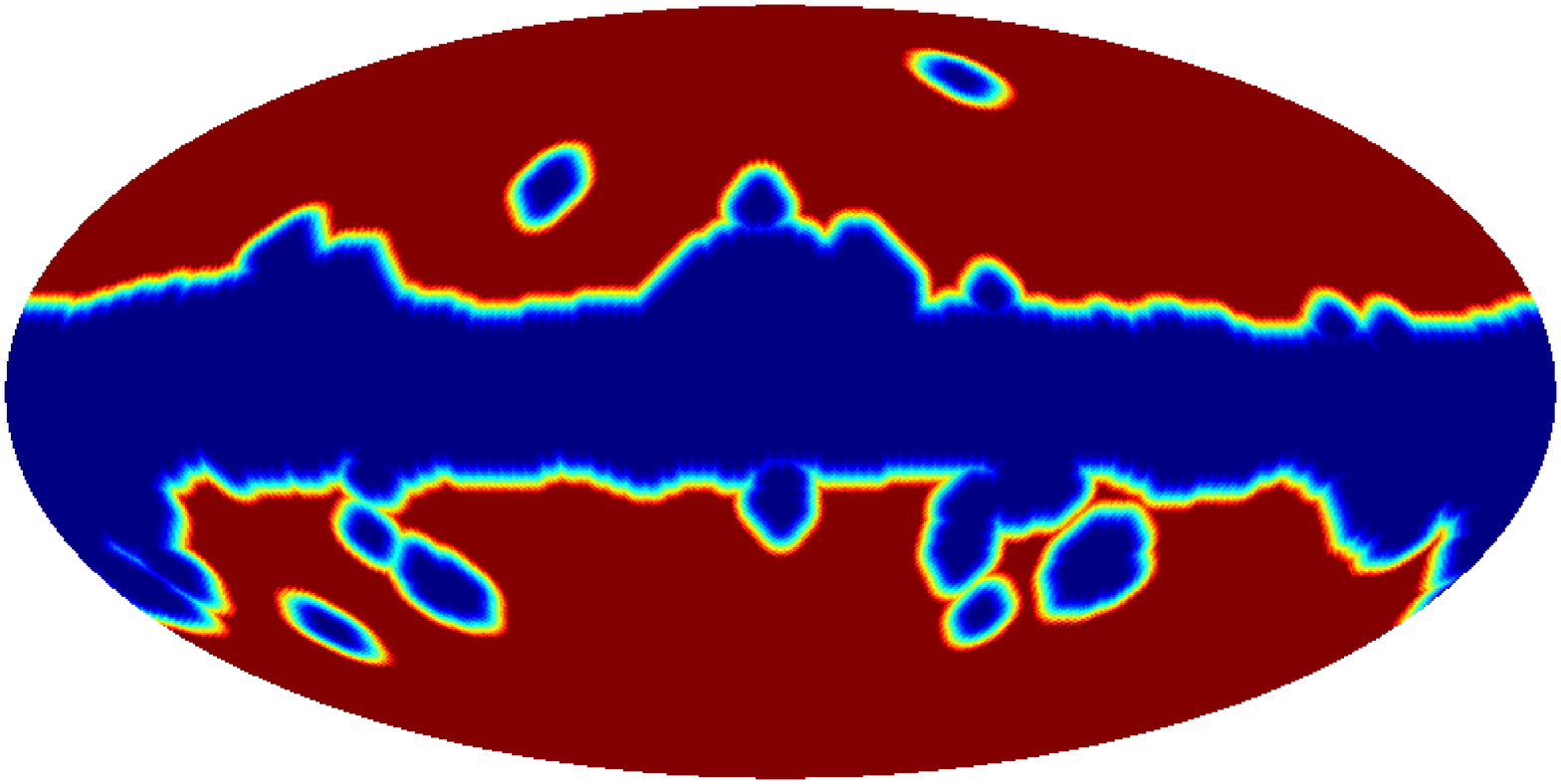}
\includegraphics[width=6cm]{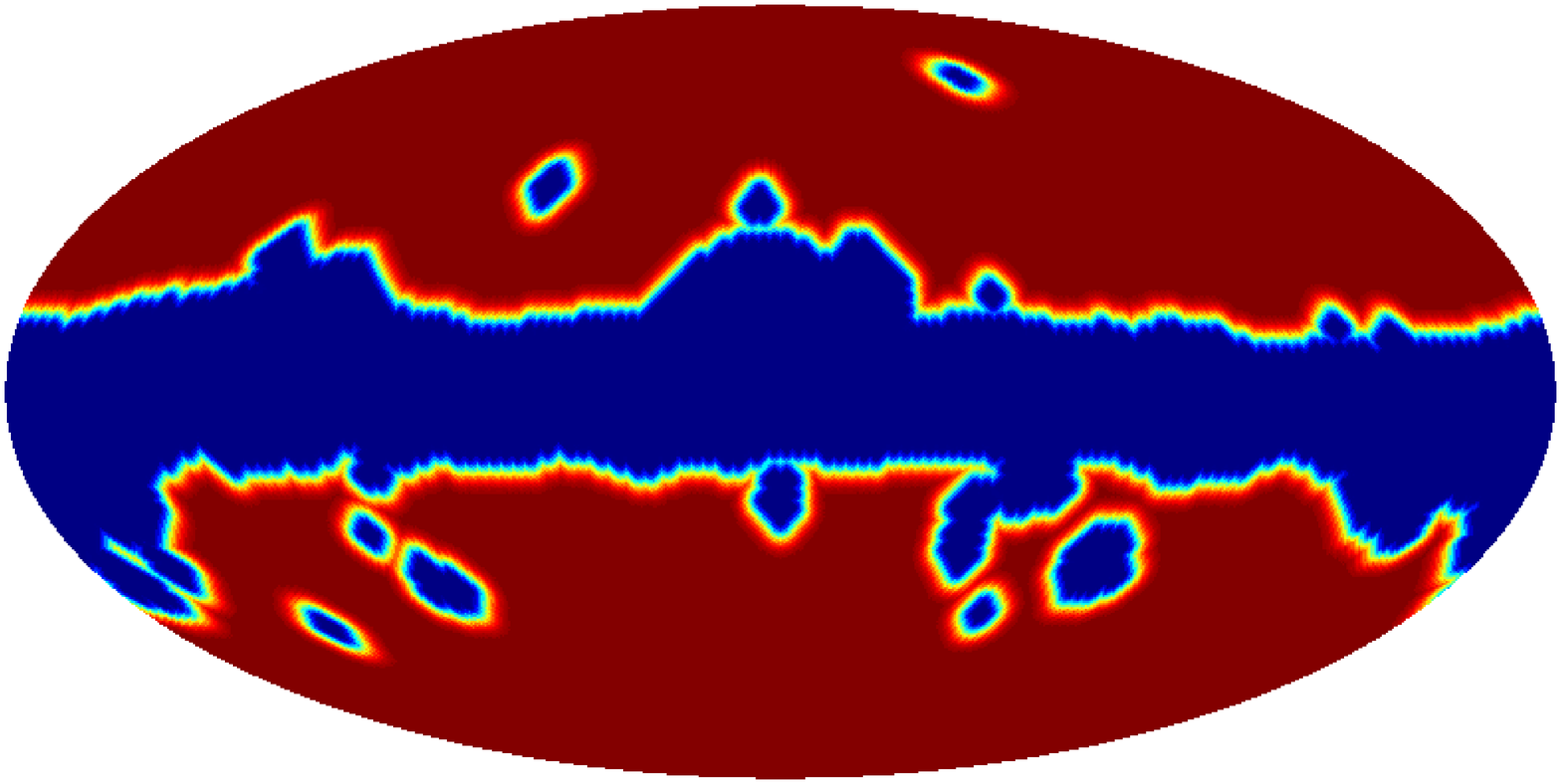}
\caption{\label{mask_ap_fig} Original sample mask (top left), cosine apodized mask (top  right), Gaussian apodized mask (bottom). The apodization angle for both cases is $7^\circ$.
}
\end{figure}

It is also useful to check the mask apodization result visually on a more complex mask. For this we combine the SMICA, SEVEM, and NILC masks from Planck \cite{Collaboration:2013vx} to obtain a sample mask (this is very similar to the Planck U73 mask). We then downgrade this mask to \code{HEALPix} $N_{side}=128$ using the \func{ud\_grade} routine and then keep only those pixels unmasked which have a value greater than $0.5$. We use this resulting mask to further test our apodization routines. The original mask, as well as $7^\circ$ apodized masks with cosine and Gaussian apodization are shown in Figure \ref{mask_ap_fig}. We have chosen the low resolution and the large apodization angle for demonstration purposes; in practice these routines are used for higher resolution maps with much smaller apodization angles.

\subsection{CMB Power Spectra}

The unit test \func{TestCMB} performs a simple test of CMB power spectra calculation. The standard cosmological parameters are set, after which the \func{CMB} module is used to calculate the temperature-temperature power spectrum. The results are compared to values obtained from running the \code{CLASS} code by itself.

\subsection{Planck and WMAP Likelihood}

The unit tests \func{TestPlanckLike} and \func{TestWMAP9Like} perform simple tests of the Planck and WMAP9 likelihood wrappers, respectively. The standard cosmological parameters as well as the foreground parameters (for Planck only) are set, then the total likelihood value is calculated. The result is compared to the values obtained from the Planck and WMAP9 likelihood codes themselves with the same $C_l$ values.

\subsection{The CMB Gibbs Sampler}

\begin{figure}
\centering
\includegraphics[width=12cm]{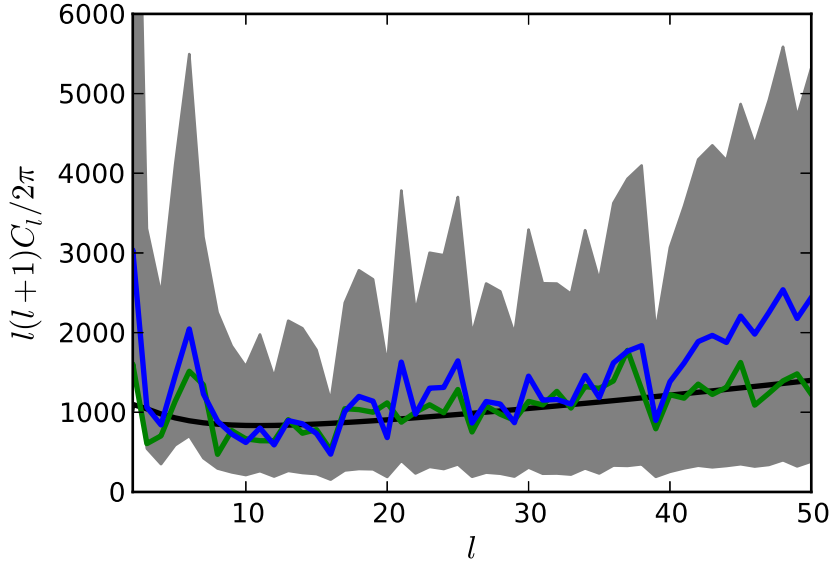}
\caption{\label{gibbs_fig} Testing the CMB Gibbs sampler. The black curve is the theoretical power spectrum, the green line is the power spectrum for the particular simulation, and the blue line is the estimate from the Gibbs sampler. The gray area is the $1\,\sigma$ confidence band.
}
\end{figure}

The class \func{TestCMBGibbs} tests the CMB Gibbs sampler. A temperature sky map is simulated with a typical power spectrum, then some white noise is added to the pixels. After masking out a part of the sky the CMB Gibbs sampler is called to create a Gibbs chain. The marginalized distribution for each $C_l$ value is then obtained from the chain and the median of the distribution is compared to the $C_l$ value of the original sky map realization.

A few subtests are performed with different sky map resolutions, noise levels, maximum $l$ values, and different sky masks. We show the result of one of the subtests in Figure \ref{gibbs_fig}. The black curve shows the theoretical power spectrum, the green curve is the power spectrum of the particular sky map realization, and the blue curve is the estimate obtained from the Gibbs chain. The gray area shows the $1\,\sigma$ confidence band. For this case we have $N_{side} = 32$ and $l_\mathrm{max} = 50$ for the Gibbs sampler. Each pixel has a white noise of $2\,\mu K$, and the map is masked using the combination of the SMICA, SEVEM, and NILC masks from Planck (see Section \ref{mask_ap_test_sec} for more details about this mask). The length of the chain is $1,000$. The estimated power spectrum from the Gibbs chain is obtained by taking the median of the marginalized distribution for each $C_l$ value.

\subsection{Simulations and Likelihood Calculation}

\begin{figure}
\centering
\includegraphics[width=12cm]{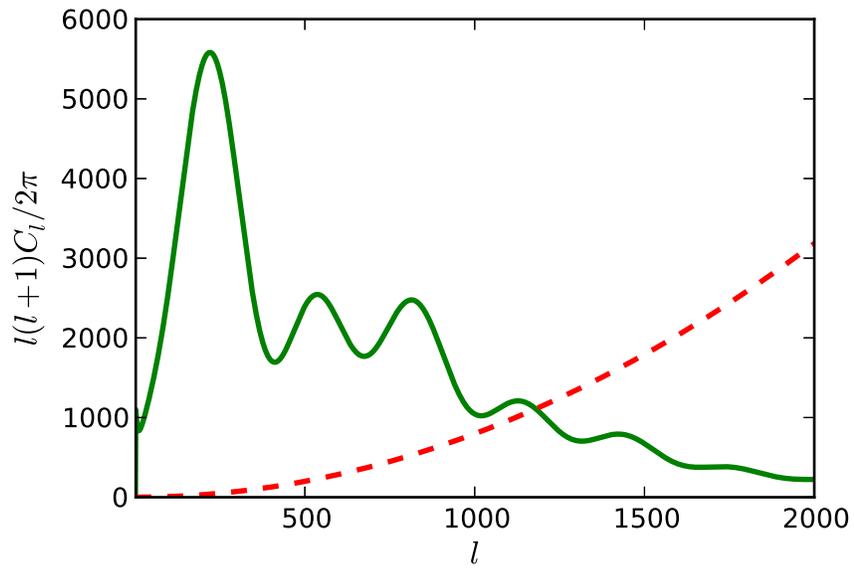}
\caption{\label{cl_fig} The signal (green solid curve) and noise (red dashed curve) power spectra for the simulations. The units are $\mu \mathrm{K}^2$.
}
\end{figure}

The classes \func{TestLikeLow} and \func{TestLikeHigh} test the low-$l$ and high-$l$ likelihood calculation routines, respectively, with simulated maps. They simulate $5,000$ sky and noise maps, mask them, then calculate the likelihoods for the signal and noise power spectra used for simulations. The simulations are done in harmonic space, then converted into pixel space using \code{HEALPix}. The signal and noise power spectra used for the simulations are shown in Figure \ref{cl_fig}. The signal power spectrum is calculated using typical $\Lambda$CDM parameter values. The maps are signal dominated up to $l\approx1200$. The simulations have \code{HEALPix} $N_{side}=2048$ and a $5^\prime$ Gaussian beam profile.

\begin{figure}
\centering
\includegraphics[width=12cm]{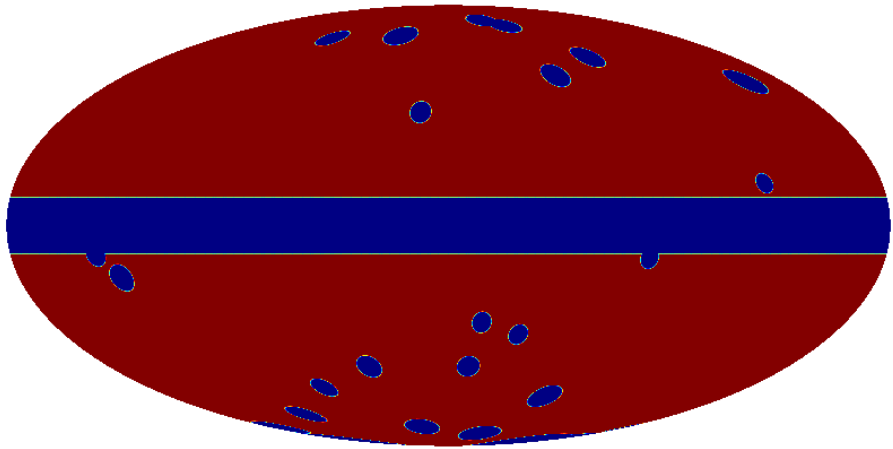}
\caption{\label{mask_highl_fig} The simulated mask used for testing the likelihood calculation routines. 
}
\end{figure}

We divide the likelihood calculation into two regimes: low-$l$ ($l=2\rightarrow30$) and high-$l$ ($l=31\rightarrow2000$). We use a simulated mask which leaves out a part of the galactic plane and some randomly selected circular regions. We apodize the mask with a $30^\prime$ cosine function. The resulting mask is shown in Figure \ref{mask_highl_fig}. We use the \func{Master} class to first calculate the best-fit power spectra, then we pass the results to the \func{LikelihoodHigh} class for calculating the likelihood values.

At low-$l$ we use the pixel space likelihood calculation class. The maps are smoothed with a $10^\circ$ Gaussian beam and downgraded to $N_{side}=16$. Since the noise level is negligible at this low resolution, we add $1\,\mu\mathrm{K}$ uniform white noise to the maps to regularize the covariance matrix inversion (see Section \ref{likelihood_sec}).

The tests finally obtain a histogram of the resulting $\chi^2$ distribution and compare them to the expected distribution, given the number of degrees of freedom. The comparison is done by simply calculating the $\chi^2$ value for the goodness of fit (not to be confused with $\chi^2$ from the likelihood code) and comparing to the expected value given the number of bins (the degrees of freedom). The tests fail if the obtained $\chi^2$ value is more than $3\,\sigma$ away from the expected value.

\begin{figure}
\centering
\includegraphics[width=6cm]{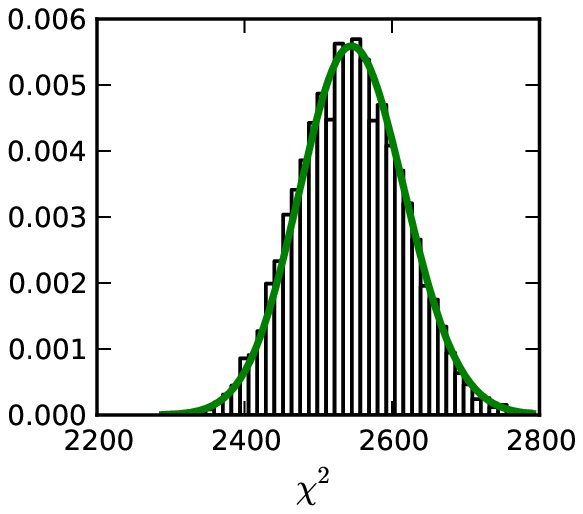}
\includegraphics[width=6cm]{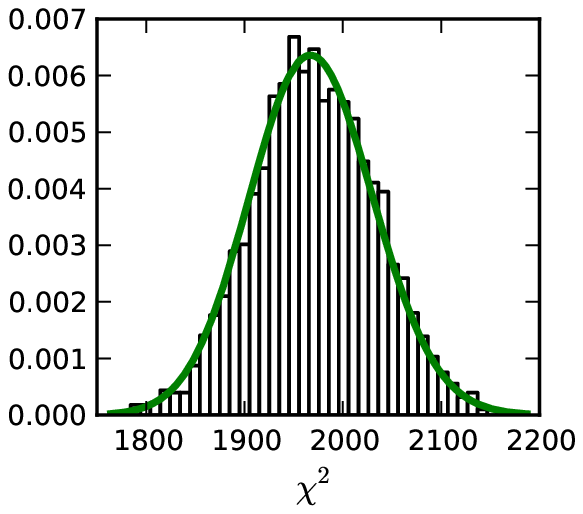}
\caption{\label{chi2_fig} The $\chi^2$ distribution from $5,000$ simulations (black histogram) compared to the expected distribution (green solid curve). Low-$l$ pixel space likelihood calculation is on the left, high-$l$ harmonic space likelihood calculation is on the right. 
}
\end{figure}

The resulting $\chi^2$ distributions and the expected curves are shown in Figure \ref{chi2_fig} for one particular run of the tests. For the low-$l$ case the expected distribution has $2,548$ degrees of freedom (the number of unmasked pixels). For the high-$l$ case the expected distribution has $1,970$ degrees of freedom (the number of $l$ values included). The $\chi^2$ value for the goodness of fit for the low-$l$ case is $37.1$ for $42$ degrees of freedom. For the high-$l$ case the goodness of fit $\chi^2$ value is $54.8$ for $41$ degrees of freedom. As we can see, the agreement between the resulting distribution and the expected distribution is excellent for both cases.

\subsection{Parameter Constraints from Planck}

\begin{table}
\renewcommand{\arraystretch}{1.5}
\setlength{\arraycolsep}{5pt}
\begin{eqnarray*}
\begin{array}{c|ccc}
  \text{Parameter} & \text{Planck} & \text{\code{Cosmo++} MH} & \text{\code{Cosmo++} MultiNest} \\
\hline
h & 0.673\pm0.012 & 0.679\pm0.011 & 0.679\pm0.012 \\
\Omega_bh^2 & 0.02205\pm0.00028 & 0.02209\pm0.00027 & 0.02212\pm0.00028 \\
\Omega_ch^2 & 0.1199\pm0.0027 & 0.1197\pm0.0025 & 0.1197\pm0.0027 \\
\tau & 0.089^{+0.012}_{-0.014} & 0.089\pm0.014 & 0.089\pm0.013 \\
n_s & 0.9603\pm0.0073 & 0.9606\pm0.0069 & 0.9601\pm0.0072 \\
\ln(10^{10}A_s) & 3.089^{+0.024}_{-0.027} & 3.088\pm0.028 & 3.088\pm0.026 \\
\end{array}
\end{eqnarray*}
\caption{Parameter constraints by Planck \cite{Collaboration:2013uv} (first column), as well as our results from \code{Cosmo++} using the Planck likelihood code together with the Metropolis-Hastings (second column) and the MultiNest sampler (third column).}
\label{param_constraints}
\end{table}

The final two tests use our parameter space samplers together with the Planck likelihood code to obtain the posterior distributions on the cosmological parameters and compare to the results published by the Planck collaboration. The class \func{TestMCMCPlanck} uses the Metropolis-Hastings sampler, and \func{TestMultinestPlanck} uses the MultiNest sampler. Both of the tests calculate the $1\,\sigma$ confidence regions on the parameters and compare them to the results by Planck. The tests fail if the medians fall more than $1\,\sigma$ away from the expected values or if the widths of the $1\,\sigma$ regions differ from the expected widths by more than $25\%$. We use the Planck likelihood code together with WMAP polarization and compare the results to those in the first column of Table 5 in \cite{Collaboration:2013uv}. Our results, together with the results released by Planck, are given in Table \ref{param_constraints}. The agreement with Planck is excellent for both of the samplers. The main noticeable difference is for the parameter $h$ which arises because we used $h$ instead of $\theta_{MC}$ (the ratio of the angular diameter distance to the last scattering surface sound horizon) for sampling. For all of the other parameters our medians do not differ from the Planck published ones by more than $0.15\,\sigma$. The widths of the $1\,\sigma$ regions do not differ by more than $8\%$. These small differences could arise because of using a different parameter space sampler or a different implementation of the same sampler.

\section{Summary}\label{summary_sec}

We have described a new numerical library for cosmology written entirely in C++. The object-oriented design has made it possible to clearly separate different parts of the library into different classes, each one of which can be used on its own. Multiple rigorous tests have been performed to check the functionality of the library. Although the code is fully documented, we have included a few examples to help the user get started.

The library is complete in the sense that it can be applied directly to publicly available data to calculate the posterior distributions and the confidence intervals for cosmological parameters. It also includes full functionality for performing CMB sky simulations, as well as numerous additional tools that are frequently used in cosmological research. However, as discussed throughout the paper, there are multiple useful features that can be added to the library. We are planning to add a few new features to the future releases of the library. The input of the scientific community will be crucial in determining the most important new features that we need to add to \code{Cosmo++}.

\vspace{0.1in}

I thank Layne Price and Richard Easther for useful discussions and comments on the manuscript. I also thank Amit Yadav for helpful discussions on the implementation of certain parts of the code.

The author wishes to acknowledge the contribution of the NeSI high performance computing facilities and the staff at the Centre for eResearch at the University of Auckland. New Zealand's national facilities are provided by the New Zealand eScience Infrastructure (NeSI) and funded jointly by NeSI's collaborator institutions and through the Ministry of Business, Innovation and Employment's Infrastructure programme {\it http://www.nesi.org.nz}.

\appendix

\section{Notation and Units}\label{notation_sec}

\subsection{Cosmological Parameters}\label{cosmo_params_ap}

We denote the Hubble parameter by $H$; it has units of $\mathrm{km}/\mathrm{s}/\mathrm{Mpc}$. The unitless form $h$ is defined by $H=100\,h\,\mathrm{km}/\mathrm{s}/\mathrm{Mpc}$.

Unitless density parameters are defined by
\begin{equation}
\Omega_i=\rho_i/\rho_\mathrm{cr}
\end{equation}
where $\rho_i$ denotes the given type of density, $\rho_\mathrm{cr}$ is the critical density, i.e. the density of the universe with zero spatial curvature
\begin{equation}
\rho_\mathrm{cr}=\frac{3H^2}{8\pi G}
\end{equation}
where $G$ denotes the gravitational constant. $\Omega_b$ denotes the baryon density, $\Omega_c$ denotes the density of cold dark matter, $\Omega_m=\Omega_b+\Omega_c$ denotes the total matter density, $\Omega_\Lambda$ denotes the dark energy density, $\Omega_\gamma$ denotes the photon density, $\Omega_\nu$ denotes the density of neutrinos, $\Omega_r=\Omega_\gamma+\Omega_\nu$ denotes the total radiation density, $\Omega_K$ denotes the curvature density.

We denote by $N_\mathrm{eff}$ the number of effective degrees of freedom for relativistic particles (neutrinos for example). In the standard $\Lambda$CDM model $N_\mathrm{eff}=3.046$ \cite{dodelson}.

Non cold dark matter particles, such as massive neutrinos, are described by their number $N_\mathrm{NCDM}$, their mass $m_{\mathrm{NCDM},i}$ in $\mathrm{eV}$, and the ratio of their temperature to the photon temperature $T_{\mathrm{NCDM},i}$.

The reionization optical depth $\tau$ is used to describe reionization. $Y_\mathrm{He}$ denotes the Helium mass fraction.

\subsection{Cosmic Microwave Background Radiation Maps}\label{cmb_maps_ap}

The Cosmic Microwave Background radiation temperature $T_\mathrm{CMB}$ is given in units of $\mathrm{K}$. The default value is $T_\mathrm{CMB}=2.726\,\mathrm{K}$. The CMB anisotropy maps have units of $\mu\mathrm{K}$ by default.

We denote the CMB temperature anisotropies in direction $\mathbf{\hat{n}}$ by $T(\mathbf{\hat{n}})$, and the polarization $Q$ and $U$ modes by $Q(\mathbf{\hat{n}})$ and $U(\mathbf{\hat{n}})$, respectively. These maps can be decomposed into spherical harmonics
\begin{equation}
T(\hat{\mathbf{n}})=\sum_{lm}T_{lm}Y_{lm}(\hat{\mathbf{n}})\,,
\end{equation}
\begin{equation}
Q(\hat{\mathbf{n}})\pm iU(\hat{\mathbf{n}})=\sum_{lm}{}_{\mp2}a_{lm}\;{}_{\mp2}Y_{lm}(\hat{\mathbf{n}})
\end{equation}
where ${}_sY_{lm}$ are the spin-weighted spherical harmonics. The polarization coefficients are further decomposed into real and imaginary parts ($E$ and $B$ modes) \cite{Page:2006hz}
\begin{equation}
{}_{\pm2}a_{lm}=E_{lm}\pm iB_{lm}\,.
\end{equation}

We use the HEALPix format \cite{Gorski:2004ku} to represent the anisotropy maps in pixel space. The value of the map $\mathcal{X}$ ($T$, $Q$, or $U$) in pixel $i$ is denoted by $\mathcal{X}_i$. This is related to the underlying map $\mathcal{X}(\mathbf{\hat{n}})$ by
\begin{equation}
\mathcal{X}_i=\int\rd\mathbf{\hat{n}}\mathcal{X}(\mathbf{\hat{n}})B_i(\mathbf{\hat{n}})
\end{equation}
where $B_i$ is the beam function at pixel $i$, and is specific to the experiment. By default, we include the HEALPix pixel window functions \cite{Gorski:2004ku} in the beam function.

Usually the beam functions have the same shape for every pixel and are axially symmetric around the center of the pixel. In this case the beam function can be decomposed into spherical harmonics as follows
\begin{equation}
B_i(\mathbf{\hat{n}})=\sum_{lm}B_lY_{lm}(\mathbf{\hat{n}}_i)Y_{lm}^*(\mathbf{\hat{n}})
\end{equation}
where $\mathbf{\hat{n}}_i$ denotes the direction of pixel $i$.

\subsection{Covariance Matrices}

We denote the two-point correlation function in pixel space by
\begin{equation}\label{cov_mat_pix}
C^\mathcal{XY}_{ij}=\vev{\mathcal{X}_i\mathcal{Y}_j}\,,
\end{equation}
and in harmonic space by
\begin{equation}\label{cov_mat_harm}
M^{\mathcal{XY}}_{lml^\prime m^\prime}=\vev{\mathcal{X}_{lm}\mathcal{Y}_{l^\prime m^\prime}^*}\,.
\end{equation}
In harmonic space $\mathcal{X}$ and $\mathcal{Y}$ denote $T$, $E$, and $B$.

When statistical rotational invariance is satisfied, the covariance matrices are diagonal in harmonic space
\begin{equation}
M^{\mathcal{XY}}_{lml^\prime m^\prime}=\delta_{ll^\prime}\delta_{mm^\prime}C^{\mathcal{XY}}_{l}\,,
\end{equation}
and $C^{\mathcal{XY}}_{l}$ denotes the conventional power spectrum in harmonic space.

The covariance matrices have units of $\mu\mathrm{K}^2$, by default. When there is no ambiguity, we sometimes omit the superscripts $\mathcal{XY}$. By default, we do not include any extra factors in $C_l$, such as $l(l+1)/2\pi$, as is sometimes done in the literature.

\subsection{Primordial Perturbations}\label{primordial_pert_ap}

We use the gauge invariant curvature perturbations on uniform density hypersurfaces $\zeta(\mathbf{x})$ to describe the primordial scalar perturbations \cite{Baumann:2009tu}. The transformation to Fourier space is given by
\begin{equation}
\zeta(\mathbf{x})=\int\frac{d^3 k}{(2\pi)^3}e^{i\mathbf{k}\cdot\mathbf{x}}\,\zeta(\mathbf{k})\,.
\end{equation}

For Gaussian perturbations, all of the information is contained in the two-point function. In case of statistically homogeneous fluctuations, the two-point function takes the form
\begin{equation}
\vev{\zeta(\mathbf{k})\zeta^*(\mathbf{k^\prime})}=(2\pi)^3\delta^3(\mathbf{k}-\mathbf{k^\prime})P_\zeta(k)\,,
\end{equation}
and $P_\zeta(k)$ is called the power spectrum.

We define the dimensionless power spectrum in Fourier space
\begin{equation}
\Delta^2_\zeta(k)\equiv\frac{k^3}{2\pi^2}P_\zeta(k)\,.
\end{equation}

Similar definitions hold for the tensor perturbations $h_{ij}$ \cite{Baumann:2009tu}. In what follows, the term ``primordial power spectrum'' will refer to the dimensionless case, unless stated otherwise.

Standard inflationary theory predicts a nearly scale invariant primordial power spectrum \cite{Baumann:2009tu}. The canonical parametrization of the power spectra is as follows
\begin{equation}\label{scalar_ps}
\Delta^2_\zeta(k)=A_s\left(\frac{k}{k_*}\right)^{n_s-1+\frac{1}{2}\alpha_s\ln(k/k_*)}\,,
\end{equation}
\begin{equation}\label{tensor_ps}
\Delta^2_h(k)=A_t\left(\frac{k}{k_*}\right)^{n_t+\frac{1}{2}\alpha_t\ln(k/k_*)}
\end{equation}
where $A_s$ and $A_t$ are the scalar and tensor amplitudes, respectively; $n_s$ and $n_t$ are the spectral indices; $\alpha_s=dn_s/d\ln k$ and $\alpha_t=dn_t/d\ln k$ are the runnings of the spectral indices. $k_*$ denotes a chosen pivot scale. The units of $k$ are $\mathrm{Mpc}^{-1}$ by default.

The tensor-to-scalar ratio is defined as
\begin{equation}
r\equiv\frac{\Delta^2_h(k_*)}{\Delta^2_\zeta(k_*)}\,.
\end{equation}

\bibliography{papers_lib,citations}

\end{document}